\documentclass[aps,prl,amsmath,amssymb,amsfonts,superscriptaddress,floatfix]{revtex4} 
\pdfoutput=1
\usepackage{amsmath,amsthm,amsfonts,amssymb}
\usepackage{graphicx}

\usepackage{color}

\newtheorem{lemma}{Lemma}

\DeclareMathAlphabet{\mathpzc}{OT1}{pzc}{m}{it}

\newcommand{\tr}{{\rm tr}}
\newcommand{\be}{\begin{equation}}
\newcommand{\ee}{\end{equation}}

\newcommand{\cO}{{\cal O}}

\begin{document}

\title{Turning Gate Synthesis Errors into Incoherent Errors}

\author{Matthew B.~Hastings}

\affiliation{Station Q, Microsoft Research, Santa Barbara, CA 93106-6105, USA}
\affiliation{Quantum Architectures and Computation Group, Microsoft Research, Redmond, WA 98052, USA}

\begin{abstract}
Using error correcting codes and fault tolerant techniques, it is possible, at least in theory, to produce logical qubits
with significantly lower error rates than the underlying physical qubits.  Suppose, however, that the gates that act on these logical qubits are only approximation of the desired gate.  This can arise, for example, in synthesizing a single qubit unitary from a set of Clifford and $T$ gates; for a generic such unitary, any finite sequence of gates only approximates the desired target.
In this case, errors in the gate can add coherently
so that, roughly, the error $\epsilon$ in the unitary of each gate must scale as $\epsilon \lesssim 1/N$, where $N$
is the number of gates.  If, however, one has the option of synthesizing one of several unitaries near the desired
target, and if an average of these options is closer to the target, we give some elementary bounds showing cases
in which the errors can be made to add incoherently by averaging over
random choices, so that, roughly, one needs $\epsilon \lesssim 1/\sqrt{N}$.
We remark on one particular application to distilling magic states where this effect happens automatically in the usual circuits.
\end{abstract}
\maketitle

There are several different settings in quantum computation where the same theme occurs: there is some error which can be reduced at the cost of an overhead, either in space (number of qubits) or time (depth of quantum circuit) or both\cite{th1,th2,th3,th4}.
To give one architecture where this theme occurs three times, imagine a system of noisy physical qubits.  Then,
imagine implementing an appropriate CSS code to produce a system of logical qubits with a smaller noise and with Clifford gate operations at high accuracy; the error rate decreases as the code distance increases which requires an increasing number of physical qubits for the same number of logical qubits.  Then, taking high accuracy Clifford operations as given, if one can approximately implement a $T$ gate at sufficiently high accuracy, one can distill magic states to produce $T$ gates with higher accuracy, again at the cost of an overhead\cite{bk,mk,mk2}.  Finally, using Clifford and $T$ gates from the first two steps, one can approximate arbitrary single qubit unitaries by a sequence of these gates, up to an error $\epsilon$ that is exponentially small in the length of the sequence; this is called gate synthesis\cite{sk,synth}.

In the simplest analysis of such a scheme, one tries to make the error rate in Clifford operations negligible so that one can assume at later steps that they are exact, and then one tries to make the error rate in the $T$ gates negligible
so that in the synthesis operation one can assume that the $T$ gates are exact.
It is important to know how serious the effect of errors are, though, as one would like to incur the minimal necessary overhead.  In this paper, we do not focus on errors which cause a large change in the state, such as a single spin flip; these are the sorts of errors that are handled by the CSS code.  Instead,
we focus on errors where the system evolves under a unitary which is close to the desired one but not exactly correct, as might occur in gate synthesis.

As a toy example, consider a single qubit.  Suppose that one wishes to apply $N$ successive unitaries $\exp(i \theta \sigma_Z)$ to this qubit.  Suppose however that one instead applies $N$ unitaries $\exp(i \theta' \sigma_Z)$ with
$\theta'-\theta=\epsilon$.
Then, in order for the error in the evolution to be small, one needs $|\epsilon| \lesssim 1/N$.
Suppose instead that each of the unitaries implements $\exp(i (\theta \pm \epsilon) \sigma_Z)$, with the sign chosen $\pm$ uniformly and independently for each qubit.  Then, the net error results from a random walk and if $|\epsilon| \lesssim 1/\sqrt{N}$ the error will likely be small.
Roughly (we do not give a precise definition) we call the first kind of scaling ``coherent" and the second kind of scaling ``incoherent".  In this paper, we give some elementary bounds showing how to achieve the incoherent
scaling if one can approximate the desired unitary by an average over other unitaries.  We then explain how this would be applied in practice, by repeatedly re-running the same algorithm with different random choices of unitaries.

The effect in this toy example would occur in simulating a Hamiltonian by Trotter-Suzuki methods\cite{lloyd}; since the same unitaries are applied repeatedly to reach a time large compared to the time step, small errors in angle can add coherently.  Indeed, the toy example is an example of this method using just a single term in the Trotter-Suzuki decomposition.  
While it was noticed before\cite{unpub} in numerical studies of Trotter-Suzuki that one could obtain good results by choosing the angle of the evolution randomly at each time step to obtain the correct average angle, here
we give a general result.  Further, we propose also varying the angle from one run of the algorithm to the next; in the toy example if one chooses the same random angles on every run, then typically there is an error in expectation value (of $\sigma_y$ if the system is initialized in the $\sigma_x=+1$ state) which is $\sim \epsilon\sqrt{N}$ as opposed to the $\sim \epsilon^2 N$ error arising from averaging over angles that we show below.

\section{General Results}
For a general setting
suppose that we wish to implement a quantum circuit composed of a sequence of unitary gates $U_1, U_2, \ldots, U_N$, so
that
the circuit implements unitary transformation $U$ defined by
\be
\label{Udef}
U=U_N U_{N-1} \ldots U_1.
\ee
Now suppose that we are only able to approximate this on some quantum computer.
For some or all of the gates, we implement the gates with some error, so that rather than
performing unitary $U_i$, we instead perform some other unitary $V_i$, with $V_i \approx U_i$.  
Let
\be
\label{Vdef}
V=V_N V_{N-1} \ldots V_1.
\ee
For applications in quantum computing, we initialize that system in some initial (possibly mixed) state $\rho$, then apply the quantum circuit, and then measure some operator $M$.  
The error in expectation is ${\rm tr}(U\rho U^\dagger M)-{\rm tr}(V \rho V^\dagger M)$.  This is upper bounded by
$\tr(|U\rho U^\dagger - V\rho V^\dagger)|) \cdot \Vert M \Vert$, where $\tr(|\ldots |)$ denotes the trace norm (i.e., the sum of the singular values) and $\Vert \ldots \Vert$ denotes the operator norm (i.e., the maximum singular value).

Of course, trivially we have $\Vert V-U \Vert \leq \sum_i \Vert V_i - U_i \Vert$, which immediately implies the bound
$\tr(|U\rho U^\dagger - V\rho V^\dagger)|) \leq  2 \sum_i \Vert V_i - U_i \Vert$.  
However, we are interested in finding a better bound with incoherent error scaling.

Suppose that for each $i$, there are several unitaries, $W_{i,1},W_{i,2}, \ldots, W_{i,n(i)}$ that we can implement,
where $n(i)$ is some integer depending on $i$,  and we can take $V_i$ to be any given one of these unitaries.  That is, we have the option to choose some integer $a_i$ and then set $V_i=W_{i,a_i}$.  We are able to choose these $a_i$ independently for each $i$.  The question then is how to do this to minimize error.
The key idea here is that while it may be difficult to show that any given sequence minimizes error, it will be
easier to show that an average of the state value $V \rho V^\dagger$
over an appropriate ensemble of sequences gives a good approximation to $U \rho U^\dagger$.

Before continuing, we briefly introduce the concept of the diamond norm.  Given a linear map ${\cal E}$
on matrices,
 a natural norm is $\Vert {\cal E}\Vert_1 \equiv {\rm max}_{\sigma,\tr(|\sigma|)=1} \tr(|{\cal E}(\sigma)|)$.  The diamond norm is defined by stabilizing this norm; it is $\Vert {\cal E} \Vert_\diamond\equiv {\rm max}_{\sigma, \tr(|\sigma|)=1} \tr(|({\cal E}\otimes I)(\sigma)|)$, where
 we have tensored ${\cal E}$ with the identity channel on an auxiliary Hilbert space of sufficiently large dimension.
The diamond norm is often used as a way to estimate the difference between two such linear maps; a bound
on the diamond norm is a stronger statement than a bound on the norm $\Vert \ldots \Vert_1$.
The diamond norm provides a useful language for the following results and it is important because it helps understand that if the unitaries $W_a$ below act only on a subsystem of the full system and act trivially on the rest of the system, then
 the norm bounds can be computed on that subsystem.

\begin{lemma}
\label{onel}
Let $W_a$ be unitaries.
Suppose that there is a probability distribution $q(a)$ such that the following holds.
Let
$\sum_a q(a) W_a \equiv \overline W$.
Let
$\delta \equiv \sum_a q(a) \Vert W_a-\overline W\Vert^2$.
Let ${\cal E}$ be the quantum channel defined by ${\cal E}(\sigma)=U \sigma U$.
Let ${\cal G}$  be the quantum channel defined by ${\cal G}(\sigma)=\sum_a q(a) W_{a} \sigma W_{a}^\dagger$.
Then, 
\be
\label{firste}
\Vert {\cal E}-{\cal G} \Vert_\diamond \leq \delta+2\Vert \overline W-U\Vert.
\ee

\begin{proof}
Let ${\cal F}$ be the linear map defined by ${\cal F}(\sigma)=\overline W \sigma \overline W^\dagger$.
We have 
$\Vert {\cal E}-{\cal F} \Vert_1\leq 2 \Vert \overline W - U \Vert$.
Indeed, we have
$\Vert {\cal E}-{\cal F} \Vert_\diamond\leq 2 \Vert \overline W- U\Vert $ since
$\Vert (\overline W - U) \otimes I \Vert =\Vert \overline W- U \Vert $.

We have
\begin{eqnarray}
({\cal G}\otimes I)(\sigma)&=&\sum_a q(a) (W_{a}\otimes I) \sigma (W_{a}\otimes I)^\dagger \\ \nonumber
&=&
\sum_a q(a) ((W_{a}-\overline W)\otimes I+\overline W\otimes I) \sigma ((W_{a}-\overline W)\otimes I+\overline W\otimes I)^\dagger \\ \nonumber
&=&(\overline W\otimes I) \sigma (\overline W\otimes I)^\dagger \\ \nonumber
&&+\sum_a q(a) ((W_{a}-\overline W)\otimes I)\sigma (\overline W\otimes I)^\dagger+{\rm h.c.}  \\ \nonumber \\
&&+\sum_a q(a) ((W_{a}-\overline W)\otimes I)\sigma ((W_{a}-\overline W)\otimes I)^\dagger.
\end{eqnarray}
The second-to-last line of the above equation vanishes, while the last line is bounded in trace norm by $\delta_i \tr( |\sigma|)$ since $\Vert (W_{a}-\overline W)\otimes I \Vert=\Vert W_{a}-\overline W \Vert$.
So
$\Vert {\cal F}-{\cal G}\Vert_\diamond \leq \delta$.  By a triangle inequality, this implies
Eq.~(\ref{firste}).
\end{proof}
\end{lemma}

Let ${\cal E}_1,\ldots,{\cal E}_N$ and ${\cal G}_1,\ldots,{\cal G}_N$ be quantum channels and let
${\cal E}\circ {\cal F}$ denote composition of channels ${\cal E},{\cal F}$.
We have the following inequality:
\be
\label{compineq}
\Vert {\cal E}_N\circ {\cal E}_{N-1} \circ \ldots \circ {\cal E}_1 - {\cal G}_N \circ {\cal G}_{N-1} \circ \ldots \circ {\cal G}_1\Vert_\diamond \leq 
\sum_{i=1}^N \Vert {\cal E}_i-{\cal G}_i \Vert_{\diamond}.
\ee
This follows because ${\cal E}_N\circ {\cal E}_{N-1} \circ \ldots \circ {\cal E}_1 - {\cal G}_N \circ {\cal G}_{N-1} \circ \ldots \circ {\cal G}=({\cal E}_N-{\cal G}_N)\circ {\cal G}_{N-1} \circ \ldots \circ {\cal G}_1+{\cal E}_N \circ ({\cal E}_{N-1}-{\cal G}_{N-1}) \circ {\cal G}_{N-1} \circ \ldots {\cal G}_1+\ldots$.
Then, use a triangle inequality and use the fact that $\tr(|{\cal E}_{i}(\sigma)|) \leq \tr(|\sigma|)$ and $\tr(|{\cal G}_{i}(\sigma)|) \leq \tr(|\sigma|)$.

This immediately implies:
\begin{lemma}
\label{twol}
Suppose that for each $i=1,\ldots,N$ there is a probability distribution $q_i(a)$, for $a=1,\ldots,n(i)$, such that the following holds.
Let
$\sum_a q_i(a) W_{i,a} \equiv \overline W_i$.
Let
$\delta_i\equiv \sum_a q_i(a) \Vert W_{i,a}-\overline W_i \Vert^2$.

Assume $|\rho|=1$.  Define $U,V$ by Eqs.~(\ref{Udef},\ref{Vdef}).  Let $V_i=W_{i,a(i)}$, with $a(i)$ chosen  independently from probability distribution $q_i(a(i))$.
Let $E[\ldots]$ denote expectation value.
Then,
\be
\label{lemmaeq}
\Bigl| E[V \rho V^\dagger - U \rho U^\dagger] \Bigr| \leq
\sum_i (\delta_i
+ 2 \Vert \overline W_i - U_i \Vert) .
\ee

\begin{proof}
Let ${\cal E}_i$ be the quantum channel defined by ${\cal E}_i(\sigma)=U_i \sigma U_i^\dagger$.
Let ${\cal G}_i$ the be quantum channel defined by ${\cal G}_i(\sigma)=\sum_a q_i(a) W_{i,a} \sigma W_{i,a}^\dagger$.
Use lemma \ref{onel} to bound $\Vert {\cal E}_i-{\cal G}_i \Vert_\diamond$.
Use Eq.~(\ref{compineq}) to bound $\Vert{\cal E}_N\circ {\cal E}_{N-1} \circ \ldots \circ {\cal E}_1 - {\cal G}_N \circ {\cal G}_{N-1} \circ \ldots \circ {\cal G}_1\Vert_\diamond$.
Note that $E[V \rho V^\dagger] ={\cal G}_N \circ \ldots {\cal G}_1(\rho)$.
\end{proof}
\end{lemma}

\section{Applications}
Lemma \ref{twol} can be applied in practice as follows.  Suppose that one wishes to estimate some expectation
value ${\rm tr}(U\rho U^\dagger M)$.  This expectation value could be estimated by applying the quantum circuit $U$ to the initial state, measuring $M$, and repeating several times to improve statistics. One  obtains approximately the same result by applying quantum circuit $V$ to the initial state, measuring $M$, and repeating several times {\it randomly resampling the unitaries $V_i$ each time} to improve statistics, with an error bounded by $\Vert M \Vert$ times the error term in Eq.~(\ref{lemmaeq}).

We now estimate the error in a simple setting.  Suppose that for each $i$, either $U_i$ can be implemented exactly (i.e., we can choose $a_i$ such that $V_i=U_i$) or $U_i$ is a rotation of a single qubit by a unitary $\exp(i \theta_i \sigma_Z)$.
Suppose for those latter $i$, we have $n(i)=2$ and $W_{i,1},W_{i,2}$ are rotations $\exp(i \theta_{i,1} \sigma_Z)$ or $\exp(i \theta_{i,2} \sigma_Z)$.
Suppose we can find probabilities $q_i(1),q_i(2)$ so that
$q_i(1) \theta_{i,1}+q_i(2) \theta_{i,2}=\theta_i$.

To estimate $\delta_i$, let $\phi_{i,1}=\theta_{i,1}-\theta_i$ and $\phi_{i,2}=\theta_{i,2}-\theta_i$.
We can compute $\Vert \overline W_i-U_i \Vert$ by considering a system consisting of just a single qubit, so both
$\overline W_i$ and $U_i$ are diagonal $2$-by-$2$ matrices; we get the same result for a system of multiple qubits with the unitaries $U_i,W_i$ just acting on one qubit as we simply tensor by the identity which does not change the operator norm (indeed, this is why the results above held for the diamond norm).
We find  $\Vert \overline W_i-U_i \Vert=|q_i(1)\exp(i \phi_{i,1})+q_i(2)\exp(i \phi_{i,2})-1|=\sqrt{(q_i(1) \cos(\phi_{i,1}) +q_i(2)\cos(\phi_{i,2}) -1)^2 +(q_i(1) \sin(\phi_{i,1}) +q_i(2) \sin(\phi_{i,2}))^2}$.

Note that $1 \geq \cos(\phi)\geq 1-\phi^2/2$ so $|q_i(1) \cos(\phi_{i,1}) +q_i(2) \cos(\phi_{i,2}) -1| \leq
\frac{q_i(1) \phi_{i,1}^2+q_i(2) \phi_{i,2}^2}{2}$.
Note also that $|q_i(1) \sin(\phi_{i,1}) +q_i(2) \sin(\phi_{i,2})|\leq \cO(|\phi_{i,1}^3+\phi_{i,2}^3|)$.
Hence,
\be
\label{eqd}
\Vert \overline W_i - U_i \Vert \leq \frac{q_i(1) \phi_{i,1}^2+q_i(2) \phi_{i,2}^2}{2}+\cO(\phi_{i,1}^4+\phi_{i,2}^4).
\ee
Similarly,
\be
\delta_i \leq q_1 \phi_{i,1}^2+q_2 \phi_{i,2}^2 + \cO(\phi_{i,1}^4+\phi_{i,2}^4).
\ee
Thus, in this setting we need to take $\phi_{i,1},\phi_{i,2}\lesssim 1/\sqrt{N}$ in order to make the error small.

\section{$T$ Gates By State Injection}
Another application of the above is to implementing $T$ gates by state injection (see for example Refs.~\onlinecite{mk,mk2}).
Assume we have an ancilla qubit (the target) in the state $2^{-1/2} (\exp(i\frac{\theta}{2}) |0\rangle+\exp(-i \frac{\theta}{2}) |1\rangle)$.
In state injection, we apply a CNOT gate from another qubit (the control) to this target, and then measure the target in the $Z$ basis.
If the measurement result is $|0\rangle$, then (up to a global phase) we implement the unitary $\exp(i \frac{\theta}{2}\sigma_Z)$ on the control,
while if the measurement is $|1\rangle$, we implement $\exp(-i\frac{\theta}{2}\sigma_Z)$.  If $\theta=\pi/4$, then this gives a way to implement the $T$ gate, assuming we can implement the $S$ gate $\exp(-i\frac{\pi}{4}\sigma_Z)$: if the measurement outcome is $|1\rangle$, we follow the measurement by an $S$ gate.
This way, with probability $1/2$ we implement $\exp(i \frac{\theta}{2}\sigma_Z)$ on the control,
and with probability $1/2$ we implement $\exp(i(\frac{\pi}{4}-\frac{\theta}{2})\sigma_Z)$ on the control.
This is the situation considered above, where if this state injection is used to implement the $i$-th unitary in the circuit we have $\theta_{i,1}=\theta/2$ and $\theta_{i,2}=\pi/4-\theta/2$ and $q_i(1)=q_i(2)=1/2$ and
$\theta_i=\pi/8$.
The randomness of the outcome of state injection automatically produces the needed averaging over two different angles.

Distillation schemes\ produce ancillas in states $2^{-1/2} (\exp(i\frac{\theta}{2}) |0\rangle+\exp(-i \frac{\theta}{2}) |1\rangle)$ with $\theta\approx \pi/4$.  
There can be both random errors (so that the value of $\theta$ varies from one ancilla to another) and
systematic errors (so that the
average value of $\theta$ differs from $\pi/4$).  However, roughly speaking, the systematic errors are turned into random
errors by the random measurement outcome.

Suppose that we are able to produce some large number of such ancillas.  Let $\mu_2$ denote the average, over
ancillas, of $(\theta-\pi/4)^2$ and let $\mu_4$ denote the average of $|\theta-\pi/4|^4$.  It does not matter whether or not the angles are independent between different ancillas.  Let $\rho$ be some initial state and $U\rho U^\dagger$ be the result of some quantum circuit including $T$ gates and $\sigma$ be the result of the quantum circuit with the $T$ gates performed (approximately) with state injection, choosing a random ancilla from this ensemble each time we do state injection.  Assume that there
are $S$ different state injections.  Then, $\tr(|\rho-\sigma|) \leq S \mu_2+S \cdot \cO(\mu_4)$.
Note that $\mu_4$ is bounded by a constant times $\mu_2$ since we can assume that all angles are bounded by $\pi$.

One may also consider a more general case where the ancilla qubit may be in state $u|0\rangle+v|1\rangle$, 
with $u=\cos(\tau)\exp(i\frac{\theta}{2})$ and $v=\sin(\tau)\exp(-i\frac{\theta}{2})$.  We now give a separate treatment of this case.  The effect of the state injection protocol is to
implement the quantum channel
${\cal G}$ defined by
${\cal G}(\sigma)=\sum_i A_i \sigma A_i^\dagger$, with
$A_1=
\begin{pmatrix}u\\&v\end{pmatrix}$
and
$A_2=
\begin{pmatrix}v \exp(i\frac{\pi}{4})\\ & u\exp(-i\frac{\pi}{4})\end{pmatrix}$.
Let $\overline A = (1/2)(A_1+A_2)$, let $\Delta_1=A_1-\overline A$ and $\Delta_2=A_2-\overline A$.
Then,
\be
{\cal G}(\sigma)=2 \overline A \sigma \overline A^\dagger+\sum_i \Delta_i \sigma \Delta_i^\dagger.
\ee
Defining ${\cal E}_i(\sigma)=\exp(i \frac{\pi}{8}\sigma_Z) \sigma \exp(-i\frac{\pi}{8}\sigma_Z)$,
we have
\be
\Vert {\cal E}-{\cal G}\Vert_\diamond \leq 2\Vert \exp(i \frac{\pi}{8}\sigma_Z) -\sqrt{2} \cdot \overline A \Vert+\sum_i \Vert \Delta_i \Vert^2.
\ee
Let $\omega=\exp(i\frac{\pi}{8})$.
Note that $\overline A={\rm diag}((u+\omega^2 v)/2,(v+\omega^{-2} u)/2)$,so
\be
\label{2ndord}
\Vert \exp(i \frac{\pi}{8}\sigma_Z) - \sqrt{2} \cdot \overline A \Vert=|\omega-(u+\omega^2 v)/\sqrt{2}|.
\ee
The right-hand side of the above equation is second order in $\tau-\frac{\pi}{4}$ and $\theta-\frac{\pi}{4}$;  i.e., it is
$\cO((\tau-\frac{\pi}{4})^2)+\cO((\theta-\frac{\pi}{4})^2)+\cO((\tau-\frac{\pi}{4})(\theta-\frac{\pi}{4}))$, though it is not analytic near $\theta=\tau=\frac{\pi}{4}$.
Also, $\Vert \Delta_i \Vert^2 = (1/4) |u-v \exp(i\frac{\pi}{4})|^2$ which is second order in $\tau-\frac{\pi}{4}$ and $\theta-\frac{\pi}{4}$.
Hence, 
$\Vert {\cal E}-{\cal G}\Vert_\diamond$ is second order in $\tau-\frac{\pi}{4}$ and $\theta-\frac{\pi}{4}$.
 (We omit the exact expression to second order, which can be found by Taylor series).

\section{Discussion}
We have considered the effects of errors in gates on a quantum computation.  An elementary calculation shows that an appropriate averaging can in some cases significantly improve the scaling, so that instead of requiring error $\epsilon\lesssim 1/N$ for $N$ gates one instead requires only $\epsilon \lesssim 1/\sqrt{N}$.  This averaging can be implemented in some cases by using a different random choice of gates each time the algorithm is run.  In some cases, such as in $T$ gates, the averaging occurs automatically.  In an architecture as mentioned at the start, 
where one has errors in the $T$ gates and where one uses a sequence of such approximate $T$ gates and Clifford gates to synthesize an approximation to another unitary, 
the diamond norm error in the approximation of the $T$
gates adds to the diamond norm error in the synthesis.

{\it Acknowledgments---} I thank D. Wecker for useful discussions.

\appendix
\end{document}